# *Smart* viscoelastic and self-healing characteristics of graphene nano-gels

Purbarun Dhar [a, *], Ajay Katiyar [b, c] and Lakshmi Sirisha Maganti [b]

[a] Department of Mechanical Engineering, Indian Institute of Technology Ropar, Rupnagar–140001, India

[b] Department of Mechanical Engineering, Indian Institute of Technology Madras, Chennai–600036, India

[c] Research and Innovation Centre (DRDO), IIT Madras Research Park, Chennai–600113, India

* Corresponding author: E–mail: purbarun@iitrpr.ac.in

Phone: +91–1881–24–2173

**Abstract**

Readily synthesizable nano-graphene and poly ethylene glycol (PEG) based stable gels have been synthesized employing an easy refluxing method and exhaustive rheological and viscoelastic characterizations have been performed to understand the nature of such complex gel systems. The gels exhibit shear thinning response with pronounced yield stress values which is indicative of a microstructure where the graphene nanoflakes intercalate (possible due to the refluxing) with the polymer chains and form a pseudo spring damper network. Experimentations on the thixotropic behavior of the gels indicate that the presence of the G nanoflakes leads to immensely augmented structural stability capable of withstanding severe impact shears. Further information about the localized interactions of the G nanoflakes with the polymer chains is revealed from the amplitude and frequency sweep analyses in both linear and non-linear viscoelastic regimes. Massively enhanced cross over amplitude values are recorded and several smart effects such as enhanced elastic behavior at increasing forcing frequencies are registered. Structural resonance induced disruption of the elastic behavior is observed for the gels for a given range of frequency and the proposition of resonance has been justified mathematically. It is observed that post this resonance bandwidth; the gels are able to *self-heal* and regain their original elastic behavior back without any external intervention. More detailed information on the viscoelastic nature of the gels has been obtained from creep and recovery compliance tests and justifications for the spring damper microstructure has been obtained. Smart features such as enhanced stress relaxation behavior with increasing strain have been observed and the same explained based on the proposed microstructure. The viscoelastic response of the gels has been mathematically modeled and it has been revealed that such complex gels can be accommodated as modified Burger's viscoelastic systems with predominant elastic/ plastic behavior. The present gels show promise in microscale actuators, vibration isolation and damping in devices and prosthetics, as active fluids in automotive suspensions, controlled motion arrestors, etc.

**Keywords:** Graphene, nanocolloids, viscoelasticity, self-healing, rheology

## 1. Introduction

Among the various allotropes of carbon, the nano allotropes, carbon nanotubes (CNT), graphene (G) and fullerenes have intrigued the research community more than other forms due to several unique and harness–able transport phenomena which hold the possibility to pave the road towards futuristic advancements in civilization. However, among these, G has niched out a special position due to its vast plethora of unique properties[1] which can revolutionize future concepts of electronics, aviation, structural engineering, healthcare, water and energy[2]. Consequently, the past two decade has been witness to research endeavors employing G in a large variety of systems, ranging from nanomedicine[3, 4], structural composites[5], water purification[6], energy harvesting[7], thermal management[8], microelectronics[9] to name a few. However, several applications of such nanomaterials require the usage of the same in the form of colloidal systems such that a particular physico–chemical property of the fluid system can be tuned in order to obtain a modified property for a required application[10, 11]. Among the more important uses of such colloidal nanosystems are formulations of nano drugs, nano coating and paints, nano elastomers and composites, electro and magnetorheological fluids and so one, all of which require efficient modification of the viscous behavior of the fluid. Both G and CNTs alongside other nanoparticle based dispersions have been used widely to harness the structural uniqueness of the nanostructures in order to modify the viscous and/or viscoelastic/ viscoplastic transport characteristics of fluids[12-18]. Often, smart characteristics such as self–healing[19] and shear dependent enhancement in viscoelastic behavior[20] are observed in such complex colloidal media.

However, often such fluids are synthesized by passive methods such as normal dispersion of the nanostructures in the base fluids and hence the colloidal components are physically separate phases. Furthermore, often CNT and graphite or graphite oxide is employed in such studies. Accordingly, it is yet unclear as to what the viscoelastic behavior of a colloidal system employing G would be in case the G nanoflakes and the base fluid molecules are actively connected and have the capability to behave as a conjoint phase. Also, most studies employ Newtonian base fluids such as oils or ethylene glycol or glycerol and accordingly the resultant fluid are almost Newtonian in behavior. Hence, it becomes difficult to assess the degree to which nanostructures impart elastic behavior to the fluid phase. In the present article, easily synthesizable G–PEG nanogels have been reported. The refluxing methodology employed to prepare the gels ensures that the G intercalates with the PEG chains and this is directly observed as effects from the viscoelastic behavior of the gels. Detailed rheological and viscoelastic studies reveal that such gels are superior compared to simple colloidal systems as well as viscoplastic composites as the gels exhibit several smart effects no observed in fluidic or elastic systems. Intriguing observations such as strain hardening of the microstructure leading to enhanced stress relaxation, self-healing with negligible elastoviscous hysteresis after adverse situations, enhanced cross over amplitudes etc. have been observed and physically and mathematically explained. The viscoelastic compliance of the fluids has been modeled based on Burgers equation and it has been inferred that the aggregate properties of a Maxwell liquid and a Kelvin-Voight solid is responsible for the observed smart properties of the gels. The gels with proper understanding of their unique viscoelastic behavior can find application is several allied areas.

## 2. Materials and methodologies

### 2.1. Synthesis and characterization of graphene

G nanoflakes synthesized via chemical route from natural graphite (Gr) powder have been utilized in the present article. The methodology for preparation of G involves a twin-step protocol[21] in line with the modified Hummers' method[22] as described in detail[23] and the process essentially yields few-layered G nanoflakes. The synthesized samples have been characterized by Raman spectroscopy and High Resolution Scanning Electron Microscopy to ensure that G has been obtained. The size distribution of G flakes has been characterized using Dynamic Light Scattering (DLS). The Raman spectra, SEM image and DLS spectra a representative synthesized sample has been illustrated in Fig. 1. Analysis of the Raman spectrum (Fig. 1 (b)) shows sharp, high intensity peaks at ~ 1350 cm$^{-1}$ and ~ 1560 cm$^{-1}$, which are the characteristic D and G bands of graphene systems[24]. The G band represents the planar stretching of the sp$^2$ hybridized carbon atoms whereas the D band manifests due to the surface defects and wrinkles caused by the exfoliation process. The broad peak at ~ 2800 cm$^{-1}$ (the 2D band) is also a characteristic of such systems and conclusively indicates the presence of G. The ratios of the intensities of 2D and G bands can be utilized to estimate the number of layers in the G samples. This ratio has been observed to be ~ 0.3–0.5, implying that the population of G is on average 3–5 sheets thick[25].

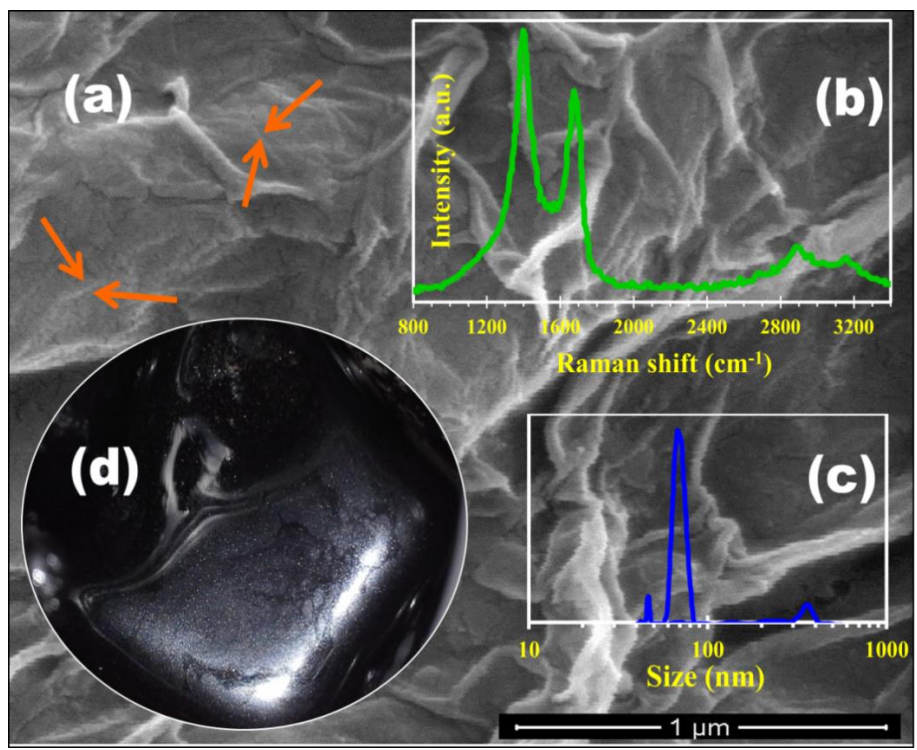

**Figure 1:** Characterization of the G samples (a) Illustrates the HRSEM image of the synthesized G. The arrows shows the nanoscale wrinkles and folds on the sheets which indicates that Gr has been exfoliated to form G (b) Raman spectroscopy intensity vs. shift plot for the G samples. (c) DLS analysis of a representative sample reveals the polydisperse size distribution with majority of sheets in the range of 60–90 nm range (d) Visual appearance of the synthesized nano gel. The physical consistency of the gel is clearly observable.

The G sample has also been characterized via HRSEM imaging, which provides conclusive evidence in favor of the formation of G structures via exfoliation, based on the surface imperfections observed. The imperfections and wrinkles appear due to stretching of the $sp^2$ hybrid layer of carbon atoms, leading to nanoscale defects and crests, as indicated by arrows in Fig. 1 (a). The strong oxidizers used in the chemical process to exfoliate the Gr to form G caused such wrinkled structure. Due to non–uniform localized gradients and reaction kinetics, the exfoliation is seldom uniform spatially and this also leads to presence of voluminous amounts of surface wrinkles and sheet defects, features in general absent on Gr flakes. The observable surface wrinkling in SEM images in fact confirms the presence of exfoliated G structures. The flake size distribution for a representative G population has been obtained using DLS analysis and has been illustrated in Fig. 1 (c). The DLS analysis reveals that major fraction of the flakes of sample lie around the peak size of 70–100 nm, with minor populations around 50 nm and 300–400 nm. Therefore, the total population of G flake sizes used in the present study lies within the meso–nanoscale regime.

**2.2. Synthesis of nanogels**

The nanogels are polymeric by nature and have been synthesized utilizing DI water and polyethylene glycol (PEG) of atomic weight 400 Da (Daltons). PEG-400 has been utilized specifically in the present study since it is a readily available polymer and among one of the most utilized, highly biocompatible, non-toxic and water soluble low molecular weight polymers. Furthermore, its thermophysical and dielectric characteristics are readily available as standardized data. The protocol developed for synthesis of G and polymer based gels has been described in the following steps and also illustrated in a flowchart in Fig 2.

1. PEG-400 is mixed homogeneously with DI water in the ratio 1:1 (by volume).

2. Raw nanocolloids are prepared by dispersing G in the above and ultrasonicating the same for 2 hours. The concentration of G added is evaluated as wt. % of the PEG-400 only.

3. The nanocolloid is refluxed at 90 $^o$C for 1.5–2 hours to obtain an intermediate gel with slimy texture.

4. The intermediate is further heated in a beaker at a reduced temperature of 50–60 $^o$C. The wide cross section of the beaker helps in more uniform evaporation of the remaining water.

5. The heating is done with constant visual inspection such that the slime does not completely dry out to form solid films. The moment at which the vapor issuing from the gel reduces drastically, the heating is discontinued. At this point, the water content of the gel reduces to trace amounts, but just enough to retain the gel form. The GNG is then allowed to cool to room temperature.

The refluxing leads to gelling as it essentially involves intercalation of the G nanoflakes with the polymer chains during the continuous process of water release and absorbance. The protocol has been experimented upon without the refluxing unit and it leads to a fluid mass where the G and the polymer are structurally independent and the phases separate during rheological tests, implying that it is an unstable colloidal system.

Care is taken not to heat the gel beyond the point where the water content reduces drastically and the vapor issuance diminishes. Heating beyond this point leads to loss of the trace amounts of water molecules too, leading to complete drying of the gel. It has been observed from repeated trials that it is the trace water that preserves the texture and form of the gel for indefinite time. Excess water content also leads to fluidic texture of the gel and therefore care is taken during preparation. A representative GNG has been illustrated in Fig. 1 (d). Three GNG samples, viz. G1, G2 and G3, consisting 2 wt. %, 1 wt. % and 0.5 wt. % G were synthesized for evaluating rheological and viscoelastic characteristics. The concentration of G in the GNGs is restricted to 2 wt. % of the polymer due to material constraints. It has been observed over and over during various synthesis trials that in general, samples with concentration over and above ~ 2 wt. % fail to form gels, with odds against gel formation around 4 of 5. Instead, the polymer–water–G system when refluxed, leads to caramelization of the polymer, leading to formation of a brownish semi-fluid which no further leads to gelation. Such a phenomenon can be explained based on the uniquely high thermal transport properties of G. Given the high thermal conductivity and heat capacity of G, presence of G nanoflakes in the polymer-water system leads to enhanced absorption of heat during refluxing. The enhanced localized heating due to presence of the G nanoflakes leads to change in the physico–chemical properties of the polymer chains, leading to caramelization. As a result, the polymer chains are either convoluted or broken and the change in microstructure induces inability to form the required gel phase.

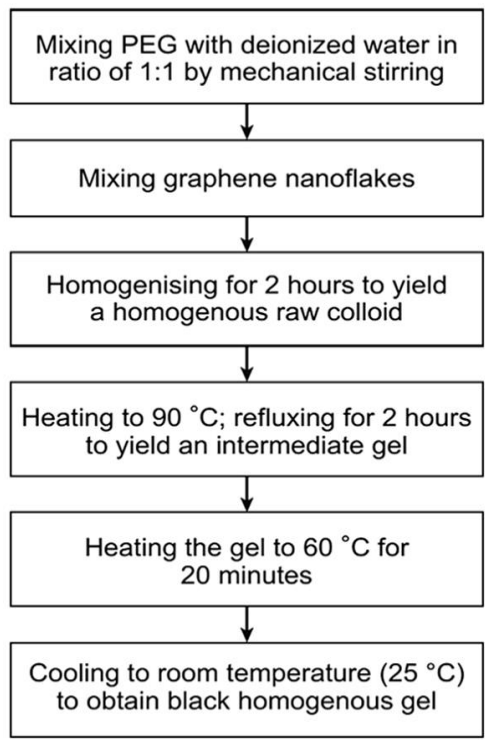

**Figure 2:** Process flowchart for the synthesis of polymer based G nanogels/pastes

## 3. Results and discussions

### 3.1. Rheological characteristics

The rheological response of the GNGs has been measured using an automated rheometer (MCR 301, Anton Paar) and the rheological data has been illustrated in Fig. 3 (a) with respect to the base PEG. It is observable that while the shear thinning essence of PEG is retained, the increasing concentration of G leads to non–linear increment in the zero shear viscosity of the gels and the GNGs do not attain Newtonian nature within the shear rate range wherein PEG attains shear independent viscosity. The presence of nanoflakes within the PEG matrix and intermingled with the polymer chains ensures that the viscosity at vanishing shear rate is higher, however, the non–linear and steep rise in the zero shear viscosity with increase in G concentration has an interesting mechanism responsible at the microscale. While it is intuitive that the zero shear viscosity would increase linearly with concentration, the interaction between graphene flakes and the PEG chains leads to enhanced viscosity. Carbon nanostructures have been reported to form pseudo–spring like structures with polymer molecules when refluxed to form fluids or gels, essentially with PEG[12]. Accordingly, from the point of view of a gel matrix with infused nanoflakes without flake–polymer interactions, the viscosity is expected to be a linear function of nanoflakes concentration. However, the linkages formed due to the nanoflakes being trapped and intercalated within the polymer matrix, leading to pseudo–spring structures, leads to additional viscous resistance due to finite elastic stiffness of the spring structures.

Under the action of external shear, the fluid matrix is set into motion, causing the linkages of the polymer–flake network to behave like microscale spring damper systems, thereby adding further resistance to the flow, leading to enhanced viscosity. Furthermore, the addition of nanomaterials leads to formation of more number of pseudo–dampers in the spring network, leading to the observed non–linear increment in viscosity. Another important and noteworthy point is the increment in the critical shear rate beyond which the gel attains shear independent viscosity. It may be observed from Fig. 3 (a) that while the base PEG attains this state at shear rate of ~ 1000 $s^{-1}$, this is delayed in the sample G3 and the samples G2 and G1 are yet to attain shear independent viscosity even at 1200 $s^{-1}$. This is yet another phenomenon which in all probability occurs due to the formation of pseudo–spring damper networks constituted by the polymer chains and the G nanoflakes. In the base polymer, the increase in shear level leads to a point at which the polymer molecules are no further able to mechanically resist the shear and hence the viscosity becomes constant with shear. The presence of such nanoscale spring–damper structures ensure that this region is largely extended and the gels are able to resist the shear effect to much larger shear values due to enhanced mechanical stiffness imposed by the nanoscale spring damper network.

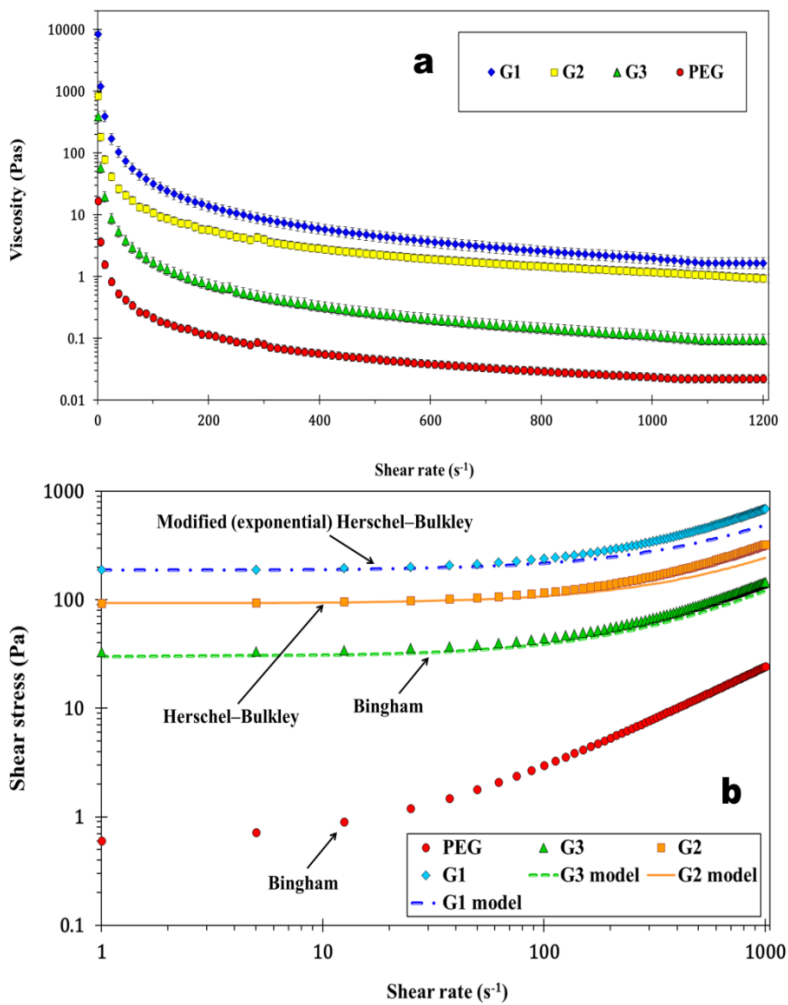

**Figure 3: (a)** Rheology of the GNG samples with respect to the base PEG illustrates that the pseudoplastic essence of the PEG is retained, however, the addition of G nanoflakes leads to enhanced thinning behavior and Newtonian behavior is delayed in terms of shear rates. The concentration of G is observed to exhibit laws of diminishing return on the zero shear viscosity. **(b)** Shear stress vs. shear rate plot for the gels reveal increment of yield stress with increasing concentration. Increasing G concentration has been found to lead to modification of the fluid from Bingham plastic to a modified form of Herschel–Bulkley fluid.

The nature of shear stress developed within the gels as a function of shear rate has been illustrated in Fig. 3(b) and it is interesting to observed that the increase in G concentration within the cells essentially leads to change in the nature of the gel itself and this again cements the spring damper microscale proposal. It is observed that the base PEG conforms to the predictions of a true Bingham fluid, whose internal stress levels ($\sigma(\gamma)$) at any imposed non–zero, finite strain rate ($\dot{\gamma}$) can be expressed in terms of the yield stress ($\sigma_y$) and the shear independent viscosity ($\mu_N$) as

$$\sigma(\dot{\gamma}) = \sigma_y + \mu_N \dot{\gamma} \qquad (1)$$

Upon addition of 0.5 wt. % G to this system and synthesizing a gel, it is observed that while the yield stress enhances by and order of magnitude, the gel retains its Bingham character, with some deviations towards higher shear rates. The addition of nanostructures leads to enhanced localized elasticity, leading to enhanced yield stress. However, it is implausible that the G flakes in the gel will shear in phase with the applied shear and it is likely that at higher shear rates, the randomness of the orientation of the G flakes in the system increases. Accordingly, a fraction of the flake population would agglomerate at high shears and resist the imposed shear with higher magnitude than the theoretically quantified value, leading to generation of higher stress levels within the microstructure and the observations mildly deviate from the Bingham theory (represented by G3 model line in Fig. 3 (b)).

On further increasing the concentration of G within the GNG to 1 wt. %, it is observed that not only does the yield stress enhance considerably, but the flow deformation behavior post yielding changes its nature from a linear Bingham system to a non–linear Herschel–Bulkley system, where the generated stress at any shear level is expressible as

$$\sigma(\dot{\gamma}) = \sigma_y + \mu_K \dot{\gamma}^n \qquad (2)$$

In Eqn. 2, the components $\mu_K$ and $n$ represent the consistency index and the power law index for the involved fluid system, respectively. Addition of larger population of G nanoflakes to the gel system leads to enhanced intercalation of the polymer chains with the nanoflakes, leading to a more robust nanoscale spring–damper system which leads to increased stiffness coefficient of the system. However, due to the presence of additional nanoflakes, the effective damping coefficient of the network also enhances, however, as the mechanisms behind stiffness and damping are different (stiffness is caused by intercalation of the polymer chains with the G nanoflakes and the damping is brought about by the nanoflakes that are not intercalated within the network), the increment in stiffness vis-à-vis damping is non–linear. This non–linearity is reflected mathematically via the consistency index. Furthermore, with increasing shear on the system, the additional resistance to the free stress distribution within the gel is hampered to a much greater extent than a gel of lower concentration due to the enhanced interflake interaction and enhanced agglomeration induced effects. Accordingly, the resistance to shear enhances non–linearly as a function of shear, which is mathematically manifested as the power law index. However, the enhanced resistance due to agglomeration and sheet collisions at higher shear leads to deviation from a constant power law index and the Hershel–Bulkley model under predicts at higher shear rates.

However, the scenario is further different when the concentration of G flakes is further increased. As observed from Fig. 3 (b), the Herschel–Bulkley model is also unable to predict the viscous characteristics of the sample G1 (2 wt. %). It is observed from the experimental data that the stress levels grow at much steeper rates than predicted by the general Hershel–Bulkley theory. Analysis reveals that while the power index can be considered constant (which reveals that the interflake agglomeration and chaotic changes in localized distribution of flakes due to shear increment), the consistency of the gel state is a function of the instantaneous imposed shear. This essentially reveals that the high concentration of G nanoflakes leading to high stiffness of the polymer chain and G flake network as well as highly non–linear damping by the non–intercalated flakes during shear. Accordingly, the internal stress levels developed within the gel microstructure enhances anomalously with increasing shear levels, which can be mathematically modeled employing a modified Hershel–Bulkley fluid model. Among the reported modified models[20], the gel sample has been observed to comply more closely with the modified exponential Herschel–Bulkley model[26] which further reveals that the stress induced within the microstructure grows exponentially with shear. The employed mathematical framework can be expressed as

$$\sigma(\dot{\gamma}) = \sigma_y(1 - e^{-m\dot{\gamma}}) + \mu_K \dot{\gamma}^n \qquad (3)$$

The stress levels within the soft structure increases exponentially with shear due to the highly inter–tangled polymer chains and nanoflakes caused by the high concentration of G nanoflakes. For the sample G1, it is found that the magnitude of the index $m$ (regularization parameter) greater than $10^3$ s leads to the best possible fit with experimental data. However, at high shear rates, the rate of growth of the stress component due to nanoflake shearing and changing localized distribution cannot be further mapped by a particular value of the index *m* and the experimental deviates to some extent from the model predictions.

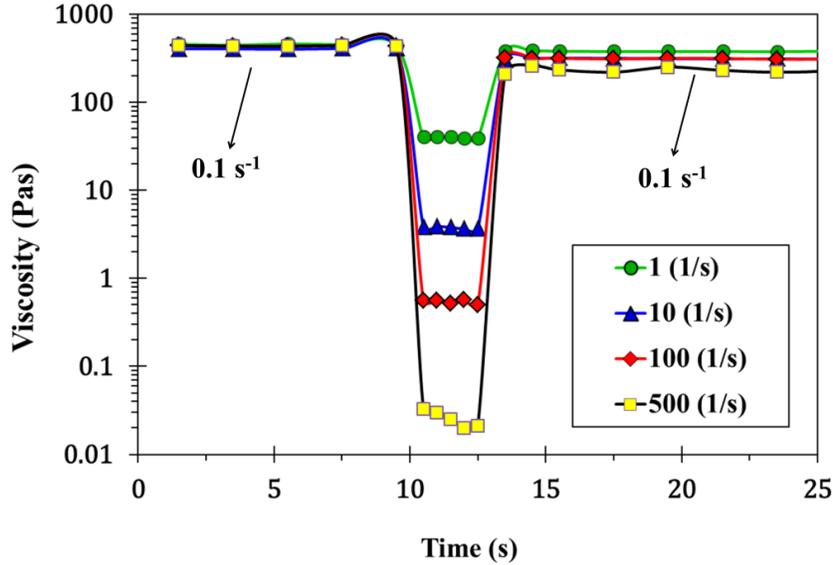

**Figure 4:** Microstructural integrity determination for the GNGs (G1, 2 wt. % sample) employing standard thixotropy test. The gel is fund to possess appreciably stable microstructure for impact shear conditions as high as 500 s$^{-1}$.

While the shear tests provide in–depth detail on the behavior of the GNGs under continuous load conditions, it is also important to understand the behavior of the gels under impact load conditions as it is pertinent from application point of view. The same can be understood via a thixotropy test and the observations from the same have been illustrated in Fig. 4. The sample G1 is subjected to a small constant shear (0.01 s$^{-1}$) for a certain period and the shear level is then suddenly increased to a desired level and maintained for a short duration. Further, the shear is decreased back to the original magnitude and maintained for certain duration such that the viscosity obtains a steady value. It is observed that for values of impact shear as high as 100 s$^{-1}$, the gel shows appreciable consistency in its microstructure, as revealed by the negligibly small hysteresis in the values of viscosity pre and post impact shear. This further provides evidence in favor of the polymer chains intercalating with the G flakes to form a pseudo spring–damper network which is able to deform freely in order to accommodate the strain generated by impact shear and then recoil back to its original form once the shear level reduced considerably. However, while the microstructure is able to withstand high shear levels, the effect of disrupted elastic nature begins playing at higher shear loads. The predominant elastic behavior of the microstructure which is able to deform and then recoil upon withdrawal of shear shows signs of pseudo plasticity in case of high shear. At 500 s$^{-1}$ shear, the viscosity during the impact phase exhibits signs of constant flow with thinning which essentially means that the microstructure yields to a certain extent. This is further confirmed from the appreciable hysteresis in the viscosity observed upon shear withdrawal which essentially implies that a minor fraction of the G-polymer chained network yields at high impacts and is unable to recover back its initial stiffness.

## 3.2. Viscoelastic characteristics

Detailed characterizations of the viscoelastic and allied response of the gels has been performed experimentally and standard and/or modified mathematical models have been employed and/or proposed in order to understand the true nature of the gels and their microstructural behavior. To begin with, the strain amplitude response of the gels at a particular frequency of applied oscillatory resistance has been considered. Figure 5 illustrates the characteristics of the gels with respect to their viscoelastic moduli under the action of sweeping amplitude at a constant oscillatory frequency. As observable from Fig 5 (a), the loss modulus of the gels shows an interesting phenomenon. Initially, beyond the short linear viscoelastic region (not shown in figure), within the non-linear zone, the storage modulus decays gradually however, the loss modulus exhibits a peak, intersects the storage modulus curve and then decays. This essentially reveals that before the crossover occurs, there is a sudden onset of the flow stresses within the sample at the microscale, leading to enhanced viscous behavior than expected. However, the more interesting part is that high frequency (500 Hz); the viscous component is able to retain a near constant value beyond the yield point, indicating that there is a mechanism of strain hardening at paly. At high frequencies, it is expected from general viscoelastic materials to show increase in the viscous characteristics post crossover, however, in the present case the same is restricted and the loss modulus value remains constant. This implies that the nanoflakes within the gel act as oscillation dampeners and at high frequencies are more active compared to low frequencies. At lower frequencies, the flakes get ample scope to align along the direction of shear motion within the gel microstructure and hence the overall fluidic behavior is able to increase with increasing strain amplitude. However, at higher frequencies, it is plausible that the gels are unable to align with the rapid oscillations and directional alignment of the flakes with the shear direction is hampered. This leads to oblique or even orthogonal alignment of a fraction of the flake population and this imparts some amount of resistance to viscous behavior, leading to the loss modulus attaining a plateau.

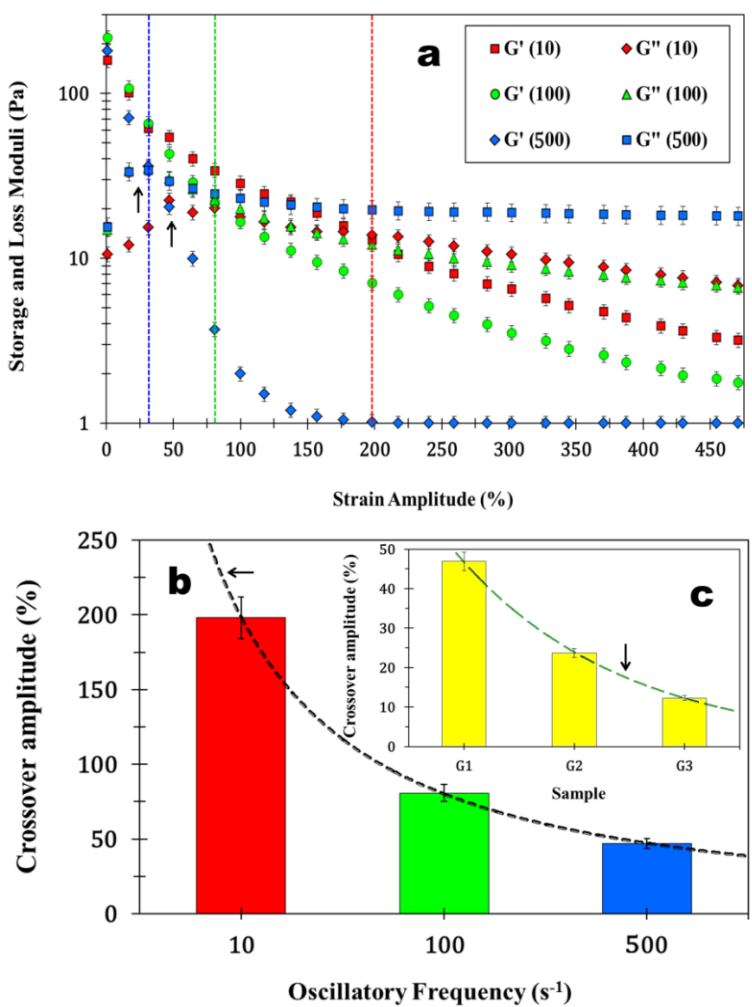

**Figure 5:** Amplitude sweep response of the GNGs illustrating various viscoelastic phenomena important for applied systems **(a)** Illustrates the behavior of the storage and loss moduli with respect to increasing imposed strain amplitude at various forcing frequencies for 2 wt. % sample

(G1). The crossover amplitudes for each case have been illustrated with a vertical dotted line. The two arrows represent the region where a transition from elastic to pseudo-plastic behavior is seen in the viscous component. **(b)** Illustrates the cross over amplitude exhibit by the sample G1 for different forcing frequencies. It may be noted that the crossover amplitude for the base PEG 400 is ~ 0.5 % at 10 Hz forcing frequency. The dotted curve represents the predictions obtained from the proposed model (Eqn. 4) **(c)** Represents the crossover amplitude of the three samples considered at 500 Hz oscillatory frequency and the dotted curve represents the predictions from the proposed model.

The crossover amplitude ($\varepsilon_{CA}$) for a particular nanogel sample with pre–determined concentration of graphene nanoflakes can be obtained by a scaled, semi empirical formulation, proposed to be of the form of

$$\varepsilon_{CA} = \left(G'\big|_{\varepsilon \to 0}\right) f^{\frac{-4\pi}{33}} \varphi \qquad (4)$$

The model is composed of three terms; where *G'* represents the magnitude of the storage modulus of the gel as the forcing amplitude tends to zero and is extracted from the plot for amplitude sweep for a required forcing frequency (*f*) and concentration of graphene (*φ*) in wt. %. For the present family of nanogels, the index of the frequency term is a constant for all concentrations of graphene and this essentially reveals that the gels are structurally consistent at the microscale; leading to behavior representable mathematically by a constant index. In the event the microstructure changes with frequency of oscillations, the whole spectrum of data would not have conformed to a single value of index for the frequency component. This is important from application perspectives as structural consistency is assured for high frequency dynamic operations as well, thereby reducing chances of failure during usage. The simplistic mathematical model is able to predict the crossover amplitude for such gels accurately with an uncertainty of ~ 10 %. The model is also accurate in predicting the cross over amplitudes for different samples, as represented by the dotted curve in Fig. 5 (c). It is interesting to note that the enhancement in cross over amplitude for the samples is a few orders of magnitude over that of PEG. This in fact is another revelation of the possible G–polymer intercalated spring–damper network system which imparts a large degree of elastic behavior to the gels. This causes the microstructure to accommodate larger degrees of oscillatory stain without undergoing pure viscous flow, leading to huge enhancements in the cross over amplitude.

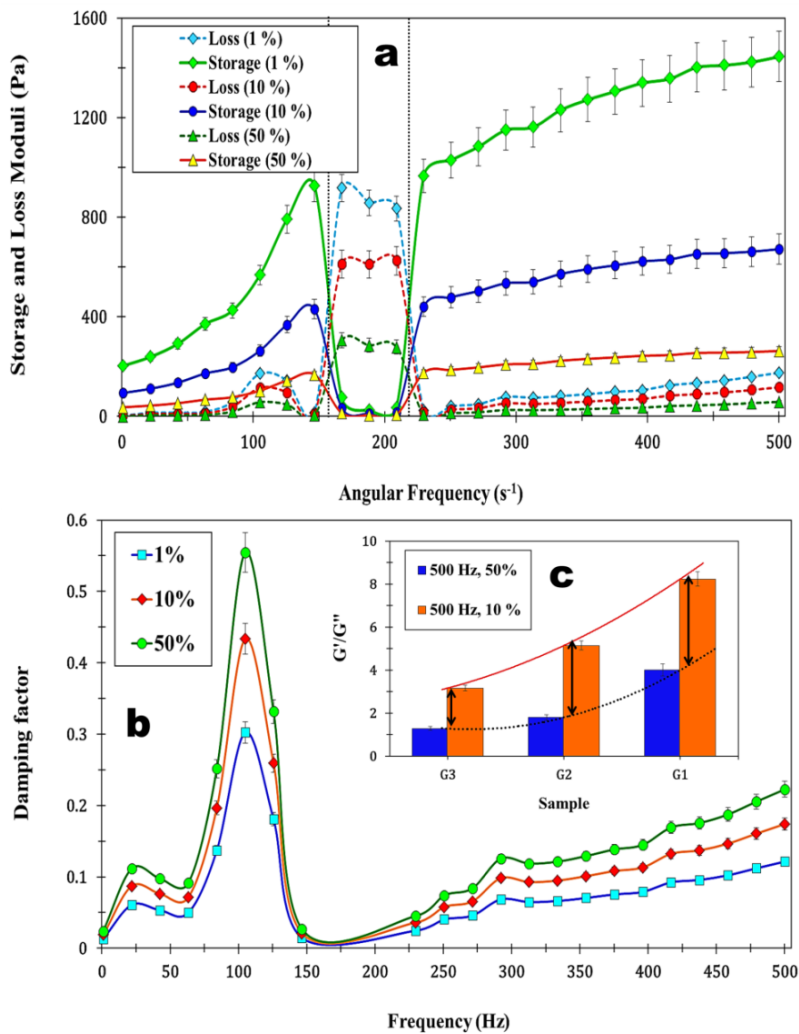

**Figure 6:** **(a)** Frequency sweep response of the GNGs in terms of storage and loss moduli exhibits an interesting region of flake oscillation resonance induced loss in elastic nature, followed by complete regain of elastic character, thus indicating at self–healing. **(b)** Illustrates the damping factor for the GNGs (2 wt. %) for different applied oscillatory strain amplitudes. **(c)** Ratio of storage to loss modulus at 500 Hz and two different levels of strain amplitude indicate the non–linear increment in elastic nature with G concentration.

Alongside the amplitude response of the viscoelastic gels, it is also important to understand the frequency response as it is often important in dynamic applications. The frequency sweep response of the gels has been illustrated in fig. 6 (a) for variant magnitudes of the imposed oscillatory strain amplitude. It is interesting to note that the gels exhibit higher magnitudes of storage moduli compared to the corresponding loss moduli for all strain amplitudes imposed. This in essence implies that the spring damper microstructure is predominantly elastic in nature under dynamic loading conditions. Furthermore, the elastic behavior of the gels is observed to increase with increasing oscillatory frequency, which brings to the forefront another smart characteristic of the gels. As the frequency of imposed oscillations increase, the oscillatory deformation rate of the polymer chain intercalated G network enhances, however, the presence of the G nanoflakes embedded within this network, which act as damping agents, are capable of damping the apparent rate of deformation. With increasing frequency, the change of direction of the deformation conjugated with the G nanoflakes leads to localized aggregation of the polymer chains and nanoflakes, leading to sort of disordered to ordered transitions within the microstructure. The presence of such aggregated ordered systems act as localized units of resistance against the high frequency of oscillatory strain and provides the gel microstructure capability to resist the deformation induced flow, leading to a form of pseudo strain hardening. With increasing frequency, the network conglomerates to additional aggregated cores within the core, leading to further enhanced elastic behavior.

Another interesting aspect revealed from the frequency sweep tests is the ability of the gels to 'self-heal' the microstructures in the event the elastic nature of the gel is lost due to working conditions and load parameters. In the present case, it has been observed consistently that the elastic modulus of the gels rapidly reduces significantly and the loss modulus enhances largely within a small frequency band ranging ~ 150–250 Hz and the phenomena has been observed to occur for all the gel samples. It is postulated that such an event occurs due to structural resonance within the embedded G nanoflakes, leading to complete disruption of the ability of the stiff nanoflakes to sustain the mechanical characteristics of the spring damper microstructure, leading to loss of elastic behavior and increment of viscous behavior by virtue of the polymer chains, now uninhibited to undergo deformation induced flow due to disruption of the G nanoflakes. The proposition of resonance induced structural failure can be provided a mathematical backing for justification. For a spring mass system constituted by a polymer chain and G nanoflake, the natural frequency would be inversely proportional to the square root of the mass of the G nanoflake as it behaves as the mass system in the present case. Considering a G nanoflake as a square with uniform density and with same number of layers, the ratio of the maximum to the minimum possible fundamental frequencies for the nanoflake population can be expressed as

$$\frac{f_1}{f_2} = \sqrt{\frac{m_{min}}{m_{max}}} = \frac{l_{min}}{l_{max}} \tag{5}$$

It can be observed from fig. 1(c) that the majority of G nanoflakes belong to the range 60–90 nm, and hence the magnitude of the ratio in Eqn. (5) is evaluated as 0.667. From fig. 6 (a), it can be seen that the ratio of the frequencies at which the proposed resonance occurs (150–250Hz) is ~ 0.6, which is fairly close to the ratio of frequency obtained if resonance were to occur within the system. Accordingly it provides a firm backing to the proposition that the observed deterioration in elastic nature within a certain frequency bandwidth is a direct result of structural resonance of the G nanoflakes. However, it is observed that beyond the frequency band, the elastic behavior is suddenly restored and the trend is the same as it would be had there be no resonance. This behavior proves the intercalated polymer G microstructure as without the presence of physical adhesion, the system would not be able to recover its initial elasticity upon termination of the resonance phase.

As far as reliability of the gels in dynamic and fluctuating load environment is concerned, deeper insight on the behavior can be obtained from the damping coefficient or loss tangent characteristics of the gels and the same has been illustrated in Fig. 6 (b). It is observed that the optimum frequency band for maximum reliability based usage of the gels is possible at 50–150 Hz, where the damping factor peaks high. This essentially signifies that the gels are sufficiently reliable for usage in basic industrial moving components as most of such components work in the frequency range of 50–100 Hz as the working frequency of several common devices is restricted by the frequency of the power supply, which are either 50 or 60 Hz. However, due to the resonance induced loss in elasticity as well as viscous nature, the gels are rendered useless in applications where the frequency of operation lies within the resonant frequency range. However, beyond this bandwidth, the gels regain their damping behavior due to the observed self-healing nature. Accordingly, the loss tangent value increases as a function of frequency (due to the proposed strain hardening like phenomena within the microstructure) and the gels can also be used reliably at higher operational frequencies. The increase in elastic nature with concentration has been illustrated in fig. 6 (c) and it has been observed that the storage modulus enhanced non–linearly as a function of G concentration. With increasing G concentration, the level of intercalation within the microstructure enhances and the stiffness of the spring damper system enhances. However, it is also true that the damping response of the system also enhances due to addition of G nanoflakes. Accordingly, each G nanoflakes has the role of inducing higher stiffness to the system as well as dampen out deformations. While the damping behavior is a function of the population of G nanoflakes, the spring stiffness is a function of its intercalation behavior with the polymer chains. As the latter is more probabilistic by nature than the former, the interplay leads to a non–linear increment in the elastic behavior and increases as intercalation density increases with addition of G nanoflakes.

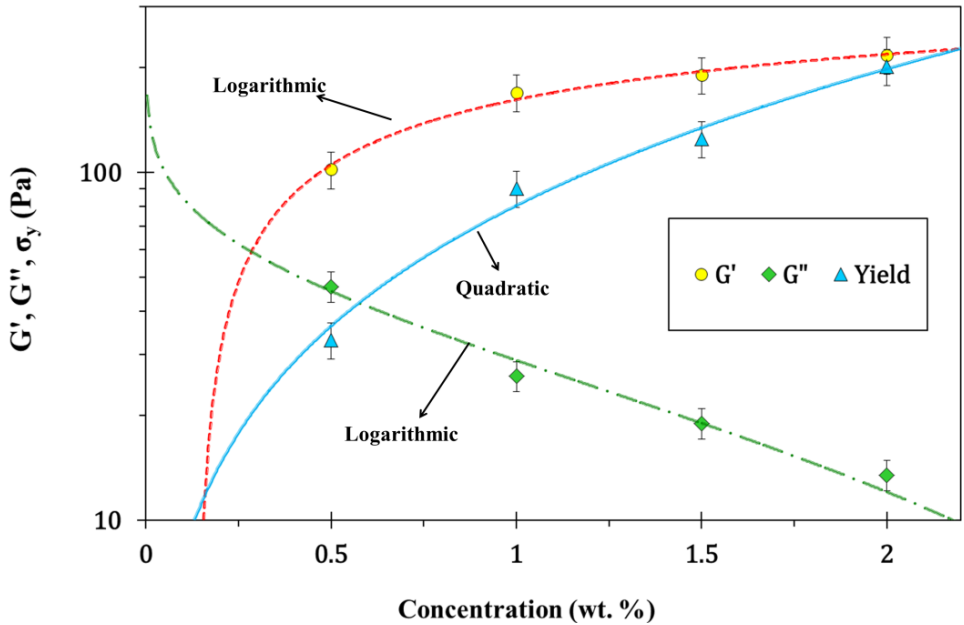

**Figure 7:** Illustration of the behavior of rheological and viscoelastic properties as function of G concentration. The storage and loss moduli obey the general logarithmic trend common to nanocolloidal gel systems[27] whereas the yield stress exhibits quadratic dependence on concentration.

An important aspect of such complex fluids is the change in their physical characteristics as a function of the dispersed phase concentration. The effect of the nanostructures on the elasticity of the complex gel has been illustrated in fig. 7. It has been observed that the storage modulus of the gel increases as a logarithmic function of the G concentration, which signifies that the rate increment decreases with increase in nanoflakes. Initially, the increment is steep and beyond a certain concentration, it begins to saturate to a plateau. Initially, with addition of the nanostructures, the intercalation of the G nanoflakes with the polymer chains during the refluxing and gelling process leads to birth of the spring–damper microstructure. This enables the gel system to significantly cease the fluid flow induced deformation under shear and the spring network is able to accommodate the deformation elastically and recoil back upon withdrawal of the external shear. This leads to a large increment in the solid modulus and a sharp deterioration in the loss modulus. With increase in nanoflakes, the number of polymer chains available for intercalation with each flake reduces, leading to nominal addition to the spring network and hence the growth rate of the apparent elasticity decreases. However, it is observed that the rate at which that the growth of the solid modulus decreases is much slower than the rate at which the loss modulus decays with increasing concentration. This indicates that the presence of the nanostructures leads to growth of elastic behavior due to intercalation with the polymer chains. However, as the number of polymer chains which intercalates per G nanoflakes is more than unity, the rate at which the polymer loses its viscous behavior is higher than the rate at which the gel attains elastic behavior. This ensures that the damping coefficient of the gels enhances as a function of concentration even when the solid modulus tends to a plateau and this is an important aspect in dynamic applications. This behavior is also reflected in the yield stress, where initially the yield stress grows with a slower pace with concentration. However, as the loss modulus deteriorates rapidly compared to the storage modulus, the caliber of the gel to yield is arrested and the yield stress grows rapidly at higher concentrations.

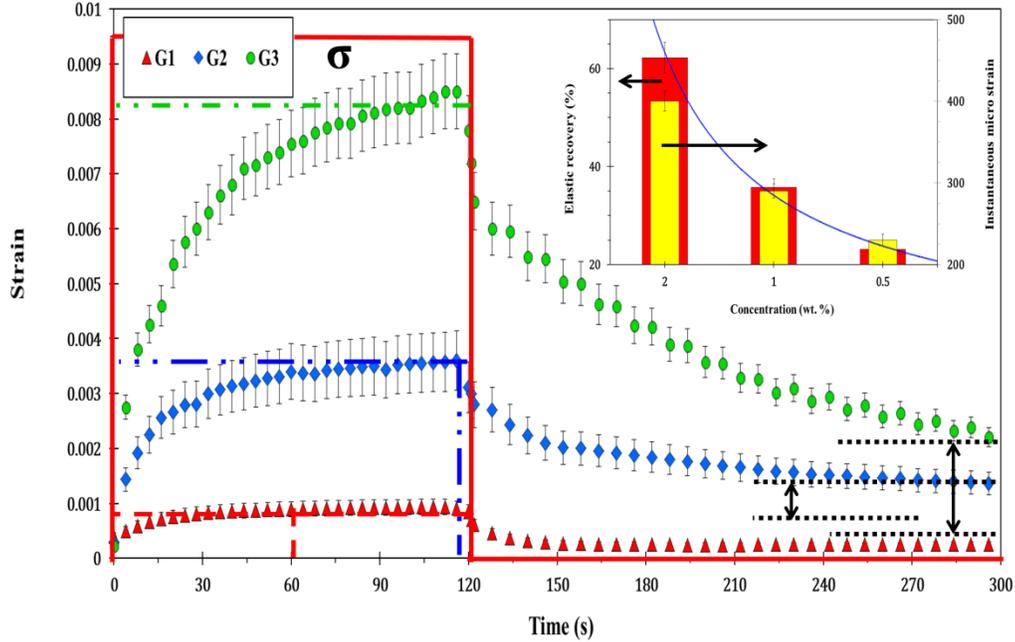

**Figure 8:** Illustration of viscoelastic creep in the GNG samples for an applied stress function (magnitude 5 Pa, thick continuous line). The strain relaxation curves show enhanced level of elastic recovery as well as the magnitude of instantaneous elastic strain with increase in G concentration (inset).

Despite the variant viscoelastic measurements, the most accurate description of the viscoelastic model which the gels comply to can be obtained from strain creep and stress relation characteristics of the gels. Fig. 8 illustrates the strain creep characteristics of the gel samples for an applied constant stress level of 5 Pa. it is observed that the concentration of G nanoflakes imparts large degree of localized elasticity to the gel microstructure, as evident from the low levels of induced net strain in the sample G1 and the large deformation in case of G3. It is further interesting to note that while in case of concentrated samples the strain attains saturated plateau after a short period of time, the low concentration system exhibits severe localized yielding and flow induced deformation, causing the strain to attain its equilibrium value at a much delayed stage. The increased nature of the localized elastic component of the microstructure with concentration can also be gauged from the magnitude of the instantaneous elastic deformation or strain levels at the moment the stress level is introduced onto the system. The inset of Fig. 8 illustrates the increasing instantaneous strain levels (right side vertical axis) with the gel samples and the non–linear increment of the initial elastic deformation with increasing concentration is evident. The mechanism can be traced back to the spring damper system microstructure, which ensures that a large deformation can be accommodated due to the elastic nature of the concentrated system, leading to large initial elastic deformation. This is similar to the concept of yield stress where the stress level accommodated within the system without flow induced deformation enhances with addition of G to the gel microstructure. The magnitude of the instantaneous elastic strain obeys a power law relationship with the G concentration (illustrated in fig. 8 inset with the curve) which further confirms the predominant power law governed behavior of the present gels.

While the creep data provides insight into the elastic and viscous components of the microstructure, important insight is also obtained from the strain relaxation behavior. After 120 s of the applied stress function, the stress is withdrawn instantaneously and the strain levels within the gels are measured and the same illustrated in Fig. 8 beyond the 120 s region. It is observed that the hysteresis in the strain levels is a function of the G concentration and decreases with increase in concentration. The sample G1 exhibits minimal hysteresis in the strain level and attains the same at the earliest after the stress is withdrawn whereas the sample G3 exhibits the largest level of hysteresis and fails to recover to the equilibrium strain level even after 300 s time frame. The elastic response of the microstructure and the ability of the G flakes within the intercalated network to behave as strain dampers can be quantified from the magnitude of the instantaneous elastic recovery. The sharp decrease in the induced strain upon immediate removal of the imparted stress constitutes the elastic recovery regime of the total strain recovery domain. The percentage of the net recovery which corresponds to the region of elastic recovery for each gel sample has been illustrated in the inset (left hand vertical axis) of fig. 8. It is noteworthy that while for a sample with 2 wt. % G the percentage of elastic recovery is ~ 60 %, the same drops to ~ 20 % for the sample with 0.5 wt. % G concentration. With increased G population, the plastic deformation within the sample reduces and the gel is able to recover a large portion of the deformation immediately. At low concentrations, while the gel exhibits enhancement in elastic recovery over the base polymer, the strain remanence within the gel is considerable and a finite amount of strain remains unrecovered even after infinite recovery period due to permanent flow stress induced restructuring within the microstructure.

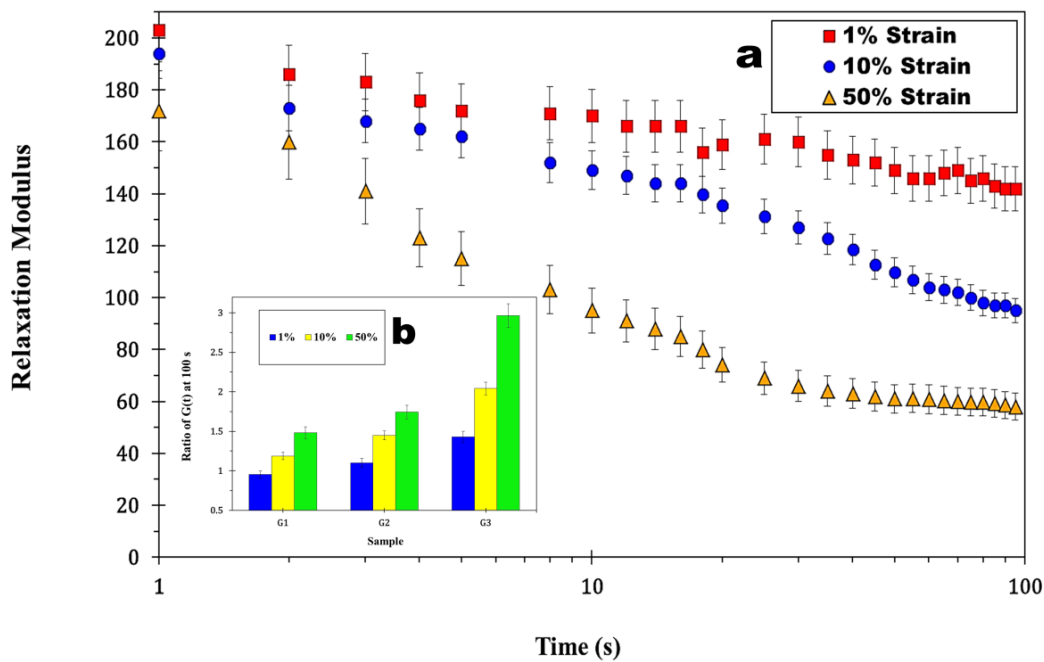

**Figure 9:** Stress relaxation characteristics of the GNGs, in terms of the relaxation modulus G(t), for applied constant strain. **(a)** Illustrates the stress relaxation for the sample G3 (2 wt. % G) over time at different applied strain levels. **(b)** The effect of G concentration and applied strain on the ratio of G(t) at t=0 to the G(t) at t=100 s, i.e. the relaxivity of the gel over time.

In addition to strain creep and recovery, the ability of the gel microstructure to relax off the induced stress levels over time also sheds insight onto the viscoelastic behavior of the intercalated G– polymer matrix. Figure 9 illustrates the stress relaxation behavior of the sample G1 under the effect of various magnitudes of imparted strains. While the G concentration is same, implying that the microstructure has the same stiffness and damping coefficients for all the cases, interesting strain dependency is observed. When the imparted strain is small in magnitude, the stress levels induced within the gels is observed to be relaxed over time. However, with increasing strain levels, it is intuitive that the induced stress is higher in magnitude and would lead to reduced stress relaxation behavior. However, the converse is observed and the gel relaxes the stress levels to a much augmented extent compared to small strain cases. This can be postulated to occur due to the intercalated polymer G network. At small strain levels, the extensional strain generated within the microstructure is minute and hence the restoring force generated is minute, in similitude to a spring mass damper system. At higher strain levels, the extension on the network is large and the larger deformation leads to enhanced restoring forces, leading to higher levels of relaxation of the induced initial stress. Such behavior is solely possible due to the formation of the compliant pseudo spring damper network due to the intercalation of polymer chains and G, thus leading to such smart characteristics. This is further backed by fig. 9 (b), which illustrates the enhanced effects of additional G concentration on the stress relaxation behavior. With increasing G, the effective stiffness of the intercalated network enhances and the restoring stress generated as response to the strain is also larger in magnitude, thus leading to higher relaxation of the stress levels. Such important properties make the gels suitable candidates in dynamic and impact load applications.

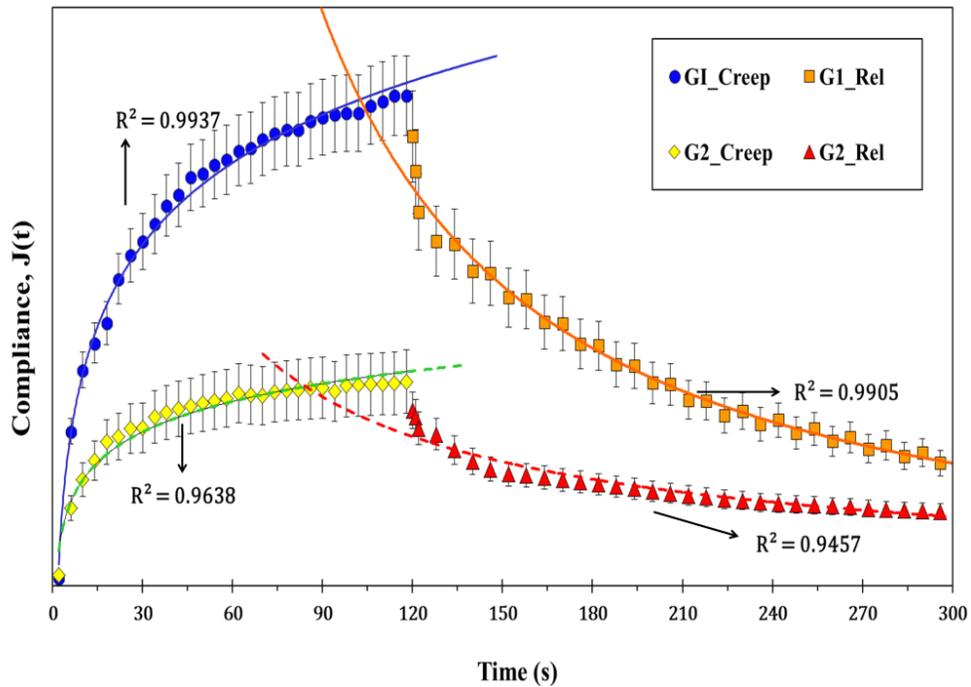

**Figure 10:** Illustration of the creep and relaxation compliances and the associated curve fitting exercise to pin point the governing viscoelastic model for the gel samples.

In order to map the exact nature of the viscoelasticity in such gel systems and their response to material properties such as G concentration, the creep and recovery behavior required to be fit to specific viscoelastic models and a best fit system needs to be obtained. Fig.

10 illustrates the fitting exercise for samples G1 and G2 in terms of the creep compliance ($J(t)$) and the recoverable compliances ($J_R(t)$), which are determined from the creep ($\gamma(t)$) and recoverable strain ($\gamma_R(t)$) as

$$J(t) = \gamma(t)/\sigma \tag{6}$$

$$J_R(t) = (\gamma_{R,init}(t) - \gamma_R(t))/\sigma \tag{7}$$

Initially, physical reasoning for the choice of a proper viscoelastic model is required to accurately trace the creep and recovery behavior of the gels. A Maxwell model maybe examined, where the viscoelastic behavior is modeled based on a spring damper system connected in series. A Maxwell fluid thus during elongation is expected to exhibit a sharp increase in the creep compliance with subsequent linear increase towards the equilibrium value. Upon withdrawal of the stress, the damper ensures instantaneous decrease in the recoverable compliance and very limited trace of gradual recovery. However, the present gels do not behave as such but exhibit some components of Maxwell fluid behavior. Similarly, a Kelvin–Voigt system can be conceived for the gels. In such a system, the creep compliance is expected to attain saturated plateau after a finite time and neither the instantaneous elastic recoverable compliance nor the residual or hysteresis recovery compliance is expected to manifest. However, the gels under some conditions exhibit such phenomena and hence are not purely Kelvin-Voigt solids. As it has been ascertained that the viscoelastic behavior is an aggregate of the two models, the standard linear model is expected to yield response levels smaller in magnitude than the smallest among the predictions by the two models. As the gels exhibit characteristics of both systems, a combined system such as fluid obeying Burgers equation maybe conceived for the present gels, where a Maxwell fluid element is in series with a Kelvin-Voight solid and the microstructure behaves as a conglomerate of the two.

The Burgers fluid formulation evaluates the creep compliance as a function of the stressing time ($t$), the viscosity of the system evaluated from the corresponding elastic ($E$) and viscous ($\mu$) components of the gels based on Maxwell (subscripted '$M$') and the Kelvin-Voigt (subscripted '$KV$') theories as

$$J(t) = \left(\frac{t}{\mu_M} + \frac{1}{E_M}\right) + \frac{1}{E_{KV}}\left(1 - e^{\frac{-G_{KV}}{\mu_{KV}}t}\right) \qquad (8)$$

Similarly, the recoverable compliance is evaluated in terms of the termination time of creep compliance ($t_C$) as

$$J_R(t) = \frac{t_C}{\mu_M} + \frac{1}{E_{KV}}\left(1 - e^{\frac{-G_{KV}}{\mu_{KV}}t_c}\right)e^{\frac{-G_{KV}}{\mu_{KV}}(t-t_c)} \qquad (9)$$

The recovery compliance is evaluated based on time elapsed beyond the point where the imposed stress is withdrawn. The parameters are chosen and the compliance curves are obtained for the optimal value of regression coefficient at which both the creep and recovery compliance curves are satisfied and the same have been illustrated in Fig. 10. For the present exercise of choosing the values of the working variables, it is ensured that both the regression coefficient is maximized for both creep and recovery compliances and simultaneously the values of viscosity employed are the closest to the zero shear viscosity of the gels. This ensures that not only is the fitting capability of the model accurate but the model does not violate the material characteristics of the gels. It is observed from the figure that the sample G1 (with 2 wt. % G) is well traced by the usage of the Burgers model and both the experimental creep as well as recovery compliances agree well with the mathematical predictions. While the regression coefficient is > 0.99, the values of viscosities employed in the model correspond to ± 15 % of the sample's zero shear viscosity. It is observed that the sample G1 fits accurately to the Burger's model and this is justified as the high G concentration leads to high elastic behavior which disrupts viscous behaviors which might cause the system to deviate from a single model prediction. However, the lower concentration of G in sample G2 leads to considerable deviations from theory in the creep–recovery overlap region. It is seen that a single value of zero shear viscosity cannot satisfy the region of elastic recovery. It is caused by the more plastic behavior of the system, where flow induced deviation in strain recovery is a possibility. Accordingly, the system behaves like a Maxwell fluid for some time period on removal of the stress and only gains the Kelvin-Voight component after some period. Accordingly, the Burger's model fits the recovery compliance beyond the point of elastic recovery.

## 4. Conclusions

In the present article, nano-graphene and PEG 400 based stable gels have been synthesized using an easy refluxing method which ensures high stability of the gels and apparently infinitely long shelf life. These gels can be categorized as complex soft matter systems due to the presence of polymer chains and G nanoflakes which form an assembled microstructure due to the refluxing process. Three gel samples of various G concentrations were synthesized and the G concentration was restricted to 2 wt. % due to events of caramelization of the polymer during refluxing. Extensive rheological and viscoelastic characterizations have been performed to understand the nature of such complex gel systems. The gels have been observed exhibit shear thinning response with enhanced yield stress values which is indicative of a microstructure where the graphene nanoflakes intercalate (possible due to the refluxing) with the polymer chains and form a pseudo spring damper network which is responsible for variant smart characteristics observed. Studies on the thixotropic behavior of the gels indicate that the presence of the G nanoflakes leads to immensely augmented structural stability capable of withstanding severe impact shears. This makes the gels of prime importance in applications involving jerk and varying load resistances. Further, information about the localized interactions of the G nanoflakes with the polymer chains are revealed from the amplitude and frequency sweep analyses in both linear and non-linear viscoelastic regimes. Largely enhanced cross over amplitude values are recorded and several smart effects such as enhanced elastic behavior at increasing forcing frequencies are registered. Structural resonance induced disruption of the elastic behavior is observed for the gels for a given range of frequency and the proposition of resonance has been justified mathematically. Beyond the resonance bandwidth; the gels are able to *self-heal* and regain their original elastic behavior back without any external intervention. More detailed information on the viscoelastic nature of the gels has been obtained from creep and recovery compliance tests and the observations justify the proposed spring damper microstructure. Smart features such as enhanced stress relaxation behavior with increasing strain have been observed and the same explained based on the proposed microstructure. The viscoelastic response of the gels has been mathematically modeled and it has been revealed that such complex gels can be accommodated as modified Burger's viscoelastic systems with predominant elastic/ plastic behavior. The present gels show promise in applications such as microscale actuators, vibration isolation and damping in devices and prosthetics, as active fluids in automotive suspensions, controlled motion arrestors, etc.

**Acknowledgement**

The authors would like to thank Dr. T Nandi, Scientist F and Head, Fuel and Lubricants division of DMSRDE (DRDO) Kanpur for the rheometer facility. PD acknowledges the partial financial support from IIT Ropar.